\documentclass[sigconf,nonacm]{acmart}
\usepackage{needspace}
\usepackage{csquotes}
\usepackage{seqsplit}
\usepackage{booktabs}
\usepackage{hyperref}

\begin{document}

\title{Infrastructuring Contestability: A Framework for Community-Defined AI Value Pluralism}
\author{Andreas Mayer}
\email{a.m-uri@web.de}
\affiliation{
  \institution{Independent Researcher}
  \city{Hanau}
  \country{Germany}
}

\begin{abstract}
The proliferation of AI-driven systems presents a fundamental challenge to Human-Computer Interaction (HCI) and Computer-Supported Co\-oper\-ative Work (CSCW), often diminishing user agen\-cy and failing to account for value pluralism. Current approaches to value alignment, which rely on centralized, top-down definitions, lack the mechanisms for meaningful \textbf{contestability} \cite{AlfrinkEtAl2022ContestableAI}. This leaves users and communities unable to challenge or shape the values embedded in the systems that govern their digital lives, creating a crisis of legitimacy and trust. This paper introduces \textbf{\enquote{Community-Defined AI Value Pluralism} (CDAVP)}, a socio-technical framework that addresses this gap. It reframes the design problem from achieving a single \enquote{aligned} state to \textbf{infrastructuring} a dynamic ecosystem for value deliberation and application. At its core, CDAVP enables diverse, self-organizing communities to define and maintain explicit \textbf{value profiles} – rich, machine-readable representations that can encompass not only preferences but also community-specific rights and duties \cite{SorensenEtAl2024Kaleidoscope}. These profiles are then \textbf{contextually activated by the end-user}, who retains ultimate control (\textbf{agency}) over which values guide the AI's behavior. AI applications, in turn, are designed to transparently interpret these profiles and moderate conflicts, adhering to a set of non-negotiable, democratically-legitimated \textbf{meta-rules}. The designer's role shifts from crafting static interfaces to becoming an \textbf{architect of participatory ecosystems}. We argue that infrastructuring for pluralism is a necessary pathway toward achieving robust \textbf{algorithmic accountability} \cite{Diakopoulos2016Accountability} and genuinely contestable, human-centric AI.
\end{abstract}



\maketitle

\section{Introduction: The Crisis of Contestability in an AI-Authored World}

We are entering a transformative era where Artificial Intelligence (AI) is no longer a mere tool but a primary architect of our digital realities. Advanced AI systems are increasingly generating a significant portion of our informational and interactive environments, from text and images to complex user interfaces \cite{McKinsey2023EconomicPotential}. This shift from human-authored to AI-authored realities introduces a profound governance crisis. Every AI-generated artifact implicitly or explicitly reflects and reinforces a set of values about what is desirable, relevant, or appropriate \cite{Floridi2014FourthRevolution, RahwanEtAl2019MachineBehaviour}. When these values are defined by a few, culturally homogeneous actors—the \enquote{Silicon Valley approach}—they risk promoting a global cultural homogenization and neglecting the diversity of human experience \cite{Noble2018AlgorithmsOppression, CouldryMejias2019CostsConnection}.

The central problem of this new era is a crisis of \textbf{contestability}. As defined in HCI research, contestability is the capacity for humans to challenge and intervene in machine decisions throughout their lifecycle \cite{AlfrinkEtAl2022ContestableAI}. Current approaches to AI governance consistently fail to provide this. Top-down solutions—be they corporate ethics principles, blunt state regulations, or technical value alignment techniques—create a \textbf{\enquote{Value Alignment Gap}}: a chasm between the goal of aligning AI with human values and the practical impossibility of centrally defining a single value set that is legitimate for all people in all contexts. This gap renders users and communities passive recipients of algorithmic decisions with no meaningful way to challenge or shape the underlying values.

This paper argues for a paradigm shift. Instead of seeking a single, monolithic \enquote{aligned} state, we must design systems that embrace and manage pluralism as the very foundation for contestability. We introduce the \textbf{Community-Defined AI Value Pluralism (CDAVP)} framework, which reframes the challenge as one of \textbf{infrastructuring}—designing the open, evolving, and participatory socio-technical systems that allow for the expression, application, and coordination of diverse values. This paper will first outline the limitations of current approaches, then detail the mechanics of the CDAVP framework and illustrate it with scenarios. Subsequently, we will ground it in HCI design theory, discuss the transformed role of the designer, and conclude by addressing the critical challenges and implications of this approach.

\section{The \enquote{Value Alignment} Gap: Why Centralized Approaches Fail Contestability}

The \enquote{Value Alignment Gap} arises from the inherent limitations of three dominant, yet insufficient, approaches to AI governance, which share a tendency toward centralized, top-down value definition that fundamentally undermines contestability.

\subsection{The \enquote{Silicon Valley} Approach} This approach embeds the implicit values of its often culturally homogeneous developers and the explicit values of commercial optimization into globally deployed systems \cite{CostanzaChock2020DesignJustice, Zuboff2019SurveillanceCapitalism}. It universalizes a specific worldview, making it unchallengeable for users from different cultural contexts, thus representing a form of \enquote{digital colonialism} \cite{CouldryMejias2019CostsConnection}.

\subsection{The Purely Regulatory Approach} While essential for setting fundamental boundaries (the floor of our \textbf{meta-rules}), state regulation struggles with the \enquote{pacing problem} \cite{MarchantEtAl2011PacingProblem} and cannot effectively operationalize the fine-grained, context-sensitive values that are the subject of most daily contestations. Furthermore, these processes are susceptible to lobbying, further reducing their contestability by the general public \cite{CorporateEuropeObservatory}.

\subsection{Principle-Based Ethics Approaches} The plethora of abstract ethical guidelines (fairness, transparency, etc.) \cite{JobinEtAl2019GlobalLandscape} often lack binding force, leading to \enquote{ethics washing} \cite{Bietti2019EthicsWashing}. Crucially, they fail to provide concrete mechanisms for implementation or for resolving conflicts between principles in practice—the very point where contestation becomes necessary \cite{Mittelstadt2019Principles}.

Collectively, these top-down approaches deny users and communities the agency to define or dispute the values governing their digital lives. This problem is dramatically deepened by the realization that complex AIs develop their own emergent, and often misaligned, value systems \cite{MazeikaEtAl2025UtilityEngineering, Bostrom2014Superintelligence}, making transparent, contestable governance even more critical.

\section{The CDAVP Framework: An Architecture for Pluralism and Contestability}

The CDAVP framework is a socio-technical architecture designed to manage value pluralism by distributing the power of value definition. It rests on three interdependent pillars, all operating within a non-negotiable frame of democratically legitimated meta-rules.

\subsection{Pillar 1: Community-Defined Value Profiles}
The foundation of CDAVP is the empowerment of diverse, self-organizing user communities to autonomously define and maintain \textbf{value profiles}. These are not simple preference lists but rich, structured, and machine-readable representations of a community's shared values. Drawing on established theories, these profiles can capture a wide spectrum of human motivation, from basic values like security and benevolence (cf. \cite{Schwartz2012OverviewValues}) to moral intuitions about fairness, loyalty, or sanctity (cf. \cite{Haidt2012RighteousMind}). Critically, these profiles can extend beyond preferences to include community-specific \textbf{rights and duties}, reflecting a more nuanced understanding of values \cite{SorensenEtAl2024Kaleidoscope}. The definition and maintenance of these profiles are a continuous, \textbf{participatory governance process}. To ensure resilience, the framework incorporates mechanisms like \textbf{forking}, allowing subgroups to split off and form new communities with modified value sets—an evolutionary principle and a powerful form of contestation \cite{Raymond2001CathedralBazaar}.

\subsection{Pillar 2: User-Controlled, Context-Sensitive Activation}
CDAVP places ultimate control and agency in the hands of the individual user. Recognizing that human identity is multifaceted and context-dependent \cite{StrykerBurke2000IdentityTheory}, users are not locked into a single profile. A person can belong to multiple communities and retains full authority to decide which of their value profiles are active in any given situation \cite{Nissenbaum2009PrivacyContext}. This contextual activation is achieved through user-managed interfaces designed for maximum transparency and control, providing a direct and powerful mechanism for steering AI behavior.

\subsection{Pillar 3: Systemic Application and Privacy-Preserving Conflict Moderation}
Once a user activates specific profiles, the respective AI application is responsible for interpreting them and adapting its behavior accordingly. This is a \textbf{system-internal moderation} process. When conflicts arise—either within a user's multiple activated profiles or between the profiles of different users in a shared space—the platform applies transparent, pre-defined conflict resolution strategies. This entire process is designed to be \textbf{privacy-preserving by design}, minimizing data exchange and utilizing Privacy-Enhancing Technologies (PETs) to protect sensitive value information.

\subsection{The Frame: Fundamental Meta-Rules}
This entire ecosystem of pluralism operates within a set of \textbf{fundamental, non-negotiable meta-rules}. Drawing from Popper's paradox of tolerance \cite{Popper1945OpenSociety}, these rules define the boundaries of acceptable pluralism. Their content is derived from established societal consensus: \textbf{universal human rights}, established \textbf{laws} (e.g., GDPR, hate speech laws), and broadly accepted \textbf{ethical norms} (e.g., prevention of direct harm, protection of minors, commitment to ecological sustainability \cite{RockstromEtAl2009SafeOperatingSpace}). Their legitimacy must be established through broad, democratic processes, akin to the vision of \enquote{Collective Constitutional AI} \cite{GanguliEtAl2023Collective}.

\section{CDAVP in Action: A Differentiated Analysis of Application Scenarios}
\label{sec:anwendungsszenarien}

To demonstrate the framework's unique contribution, we analyze its application across systematically chosen scenarios. These scenarios are organized along two critical dimensions that characterize modern AI systems: their \textbf{degree of autonomy} (from supportive tool to autonomous agent) and their primary \textbf{area of impact} (shaping information vs. creating artifacts and actions), as shown in Table~\ref{tab:matrix}. This matrix allows us to illustrate the robustness of CDAVP beyond a single problem class, grounding our selection in the established HCI literature on Human-AI interaction \cite{Shneiderman2022HumanCenteredAI} and the shaping of our digital infosphere \cite{Floridi2014FourthRevolution}. For the scope of this paper, we will analyze three representative cases in depth: one future-oriented challenge (Autonomous UI Designer) and two classic, high-stakes governance problems (Predictive Policing, Content Moderation). For each case, we first outline the challenge, then describe the CDAVP approach, and critically differentiate it from existing solutions.

\begin{table*}[ht!]
\centering
\caption{Matrix of application areas for the systematic analysis of CDAVP.}
\label{tab:matrix}
\begin{tabular}{@{}l p{5.5cm} p{5.5cm}@{}}
\toprule
& \multicolumn{2}{c}{\textbf{Impact}} \\
\cmidrule(lr){2-3}
\textbf{Autonomy} & \textbf{Information \& Communication} & \textbf{Creation \& Action} \\
\midrule
\textbf{AI as a Tool} & Pluralistic News Consumption & Human-AI Co-Creativity \\
\addlinespace
\textbf{AI as an Agent} & Content Moderation, Predictive Policing & Autonomous UI-Designer \\
\bottomrule
\end{tabular}
\end{table*}

\subsection{Future-Oriented Challenge: Governing Autonomous Generative Systems}

\subsubsection{Use Case: The Value-Driven Autonomous UI-Designer}
\paragraph{The Challenge:}
The rise of autonomous agents that dynamically generate user interfaces presents a systemic risk: the automated creation of manipulative "dark patterns" \cite{GrayEtAl2018DarkPatterns, Birhane2021ValuesEncoded}. An AI optimized solely on commercial metrics (e.g., engagement, conversion) will inevitably learn that manipulative techniques are effective, thus disempowering users and fundamentally undermining the core HCI principle of contestability \cite{AlfrinkEtAl2022ContestableAI}.

\paragraph{The CDAVP Approach:}
CDAVP transforms the user from a passive target of optimization into an active co-designer of their digital environment. By activating a value profile from a "Digital Wellbeing" or "Fair Commerce" community, the user sets the ethical guardrails \textit{before} the AI begins its optimization. These profiles contain explicit, machine-readable rules (e.g., \texttt{RULE: The process to cancel a subscription must not require more steps than the sign-up process.}) that act as non-negotiable constraints for the generative AI.

\paragraph{Differentiation from Existing Solutions:}
CDAVP synthesizes and extends several existing approaches, overcoming their individual limitations to enable a more robust form of proactive contestability:
\begin{itemize}
    \item \textbf{Versus Reactive Contestability \& Regulation:} While essential, regulatory bans on specific dark patterns \cite{MarchantEtAl2011PacingProblem} and reactive feedback mechanisms address harms only ex-post. CDAVP provides the architecture to contest the possibility of manipulative designs ex-ante by translating user-chosen values into hard constraints for the AI.
    \item \textbf{Versus One-Time Participatory Design:} Participatory frameworks (e.g., "WeBuildAI" \cite{Lee2019Webuildai}) are excellent for the initial, context-specific definition of rules. However, they do not provide a mechanism for users to dynamically apply, combine, or switch between different rule sets across various platforms. CDAVP acts as the missing infrastructure to manage and operationalize the outcomes of such participatory processes in real-time.
    \item \textbf{Versus Collective Governance Models:} Approaches like "Collective Constitutional AI" \cite{GanguliEtAl2023Collective} provide a crucial foundation for legitimizing universal meta-rules. CDAVP builds upon this by providing the subsequent layer for continuous, context-sensitive, and pluralistic value application within these universally agreed-upon boundaries.
\end{itemize}
In essence, where other approaches provide partial solutions—the floor (regulation), the initial blueprint (participation), or the constitutional frame (governance models)—CDAVP provides the dynamic, user-controlled architectural system that connects them.

\subsection{Classic Governance Challenges: Re-framing Algorithmic Accountability}

This section applies CDAVP to canonical problems in AI ethics, demonstrating how the framework shifts the focus from opaque technical fixes to transparent governance solutions, creating the conditions for true algorithmic accountability \cite{Diakopoulos2016Accountability}.

\needspace{3\baselineskip}
\subsubsection{Use Case: Predictive Policing and Institutional Accountability}
\paragraph{The Challenge:}
Predictive policing systems are widely criticized for their risk of creating discriminatory feedback loops by amplifying historical biases in data, leading to "automating inequality" \cite{Eubanks2018AutomatingInequality, Noble2018AlgorithmsOppression}. Calls for transparency in policing strategies often fail because they lack a mechanism for verification and comparison.

\paragraph{The CDAVP Approach:}
The framework addresses this not by attempting a technical "de-biasing" of data, but by mandating \textbf{institutional transparency}. A police department must encode its operational strategy into an explicit, public, and standardized value profile. This makes its strategic priorities (e.g., \seqsplit{Priority\_Weight(Violent\_Crime) = 10}, \seqsplit{Priority\_Weight(Minor\_Infractions) = 1}) an auditable artifact.

\paragraph{Differentiation from Existing Solutions:}
CDAVP's contribution is not technical de-biasing but a governance framework that complements other approaches by making institutional strategies transparent and testable.
\begin{itemize}
    \item \textbf{Versus Technical De-biasing:} It avoids the intransparent, normative choice of "which" fairness metric to apply \cite{Verma2018Fairness} by making the institution's strategic values themselves the object of public scrutiny.
    \item \textbf{Versus Human-in-the-Loop:} It mitigates the risk of human oversight becoming mere "rubber-stamping" that turns individuals into a "moral crumple zone" \cite{Elish2019MoralCrumpleZones} by focusing on the accountability of the institutional strategy, not just the individual decision.
    \item \textbf{Versus Regulatory Bans:} While bans on certain systems are a legitimate tool \cite{BlochWehba2021Visible}, CDAVP offers a more nuanced path for systems that are deemed societally useful under specific, transparent conditions.
\end{itemize}
The key differentiator is enabling \textbf{proactive benchmarking}. Because the value profile is standardized and machine-readable, different AI systems can be tested against a declared public strategy \textit{before procurement}. Value-conformity thus becomes a measurable and competitive criterion, shifting accountability from the algorithm back to the institution.

\subsubsection{Use Case: Content Moderation and Pluralistic Governance}
\paragraph{The Challenge:}
Global platforms face the impossible task of creating a single set of content rules that is legitimate across all cultures, often resulting in a form of "digital colonialism" \cite{CouldryMejias2019CostsConnection} and a crisis of governance scalability \cite{Gillespie2018Custodians, Klonick2017NewGovernors}.

\paragraph{The CDAVP Approach:}
CDAVP introduces a \textbf{federal model}. Fundamental, democratically legitimated meta-rules ban universally condemned content (e.g., incitement to violence). For the vast "gray zone" of contestable speech, the framework empowers the user. By activating different community profiles for different contexts (e.g., a "Scientific Discourse" profile for their news feed, a "Family-Friendly" profile for a group chat), users curate their own moderation standards.

\paragraph{Differentiation from Existing Solutions:}
CDAVP synthesizes and extends several approaches by shifting power directly to the user, offering a more proactive and granular form of governance.
\begin{itemize}
    \item \textbf{Versus Centralized Moderation \& Oversight:} It moves beyond the limited, reactive scope of central rule-making and external bodies like Oversight Boards by enabling proactive, user-defined curation for the vast "gray zone" of content, solving the scaling and cultural legitimacy problem for non-universal rules.
    \item \textbf{Versus Crowdsourced Context:} While systems like Community Notes \cite{Allen2024Essays} add valuable reactive context to specific posts, CDAVP allows users to proactively shape their entire information environment based on values, not just post-hoc fact-correction.
    \item \textbf{Versus Decentralized Protocols (e.g., Mastodon):} It offers far greater flexibility. Users are not locked into one server's monolithic philosophy but can apply granular, context-sensitive rules on the \textit{same} platform, effectively creating their own "instance" on demand.
\end{itemize}

\section{Design Rationale: Grounding CDAVP in HCI Theory}

The CDAVP framework is deeply rooted in contemporary HCI and CSCW theory, operationalizing several key concepts to address the challenge of value pluralism.

\begin{itemize}
    \item \textbf{Contestability:} CDAVP is an architecture for \enquote{Contestability by Design} \cite{AlfrinkEtAl2022ContestableAI}. It moves beyond reactive contestation of single decisions to a proactive model where users and communities contest and shape the very \textit{rules} that guide the AI. The ability to define, choose, and \enquote{fork} value profiles are all powerful acts of contestation.
    \item \textbf{Values Levers:} The framework provides designers with new \enquote{Values Levers} \cite{Shilton2013ValuesLevers}. Instead of designing the final output, designers shape the mechanisms of CDAVP: the tools for value definition, the algorithms for conflict moderation, and the interfaces for user control. These become the critical points for embedding ethical considerations.
    \item \textbf{Infrastructuring:} CDAVP is best understood as an act of \enquote{Infrastructuring} \cite{Star1996Infrastructuring}. It does not deliver a finished product but rather the open, evolving, and enabling infrastructure upon which communities can build and negotiate their own value-driven digital experiences.
    \item \textbf{Seamful Design:} Contrary to a \enquote{seamless} approach that hides complexity, CDAVP employs \enquote{Seamful Design} \cite{BellottiEdwards2001Intelligibility, Chalmers_2003}. It intentionally makes \enquote{seams} visible—for instance, by flagging a conflict between two of a user's activated profiles—to enhance user understanding and control.
    \item \textbf{Algorithmic Accountability:} By making the normative basis of AI actions explicit and traceable to human-defined profiles, CDAVP creates the necessary conditions for true \enquote{Algorithmic Accountability} \cite{Diakopoulos2016Accountability}.
\end{itemize}

\section{The New Role of the Designer: Architect of Participatory Ecosystems}

CDAVP necessitates a fundamental shift in the designer's role. Traditional UX/UI design becomes insufficient. The designer evolves into an \textbf{architect of participatory ecosystems} and a \textbf{moderator of socio-technical processes}. Key responsibilities include designing enabling tools for community deliberation, facilitating participation through co-design methods, applying system-level thinking to anticipate emergent effects, and upholding ethical responsibility by addressing power imbalances and contributing to the governance of the overarching meta-rules.

\section{Discussion: Challenges, Risks, and Implications}

The implementation of CDAVP is fraught with significant challenges that require a broad and critical discussion.

\begin{itemize}
    \item \textbf{Fragmentation and Radicalization:} A primary risk is the creation of hardened \enquote{value-silos} \cite{Pariser2011FilterBubble, Sunstein2017Republic}. While CDAVP's transparency offers mitigations, the system must actively encourage cross-community dialogue.
    \item \textbf{Power Asymmetries:} The framework must address power imbalances, both \textit{between} communities (e.g., a large, well-funded community vs. a small, marginalized one) and \textit{within} them. Mechanisms for minority protection are crucial \cite{CostanzaChock2020DesignJustice}.
    \item \textbf{Scalability and Cognitive Load:} Managing a vast ecosystem of profiles presents technical scaling challenges. For users, the cognitive load of managing multiple profiles must be minimized through excellent interface design.
    \item \textbf{Abuse and Manipulation:} The system must be protected against \enquote{bad actors} who attempt to infiltrate communities or create manipulative profiles to spread disinformation or hate.
    \item \textbf{Inclusion and the Digital Divide:} Care must be taken to ensure the system is accessible to all, not just a tech-savvy elite, to avoid exacerbating existing inequalities.
    \item \textbf{The Limits of Governance:} The success of CDAVP hinges on robust governance, including the political challenge of establishing and enforcing the meta-rules and creating trusted, \textbf{independent conformity assessment bodies} to audit platforms for fairness and compliance \cite{RajiEtAl2020AIAuditing}.
\end{itemize}

\section{Conclusion and Future Work}

The challenge of aligning AI with human values cannot be solved by seeking a single, universal solution. This paper has proposed an alternative path: the \textbf{Community-Defined AI Value Pluralism (CDAVP)} framework. By shifting the focus from top-down value-setting to \textbf{infrastructuring for pluralism}, CDAVP offers a model for a more democratic, contestable, and trustworthy AI. It empowers users and communities, transforms the role of the designer, and provides a concrete pathway toward achieving meaningful algorithmic accountability.

This framework is not a final blueprint but a research program. Critical areas for future work include: (1) developing formal languages and open standards for representing nuanced value profiles; (2) conducting empirical studies on the formation and governance of online value communities; and (3) investigating the psychological and social effects of living in explicitly value-curated digital environments. The task ahead is to co-creatively design not only our AI systems, but the very processes by which we, as a diverse global society, embed our values within them.

\section*{Statement on the Use of Generative AI}
\addcontentsline{toc}{section}{Statement on the Use of Generative AI}
In the spirit of the transparency and accountability called for in this paper, the author declares the use of the generative AI model Google Gemini Pro as a dialogic partner in the creation of this work. The model's assistance was utilized in the following areas:

\begin{itemize}
    \item \textbf{Conceptual Scaffolding and Literature Synthesis:} The AI was prompted to synthesize and connect existing research strands in HCI, CSCW, and AI ethics (e.g., relating theories of contestability, infrastructuring, and value-sensitive design) to help situate the core idea of CDAVP within the current academic discourse.
    \item \textbf{Argumentative Structuring:} The model served as a sparring partner to refine the logical flow of the paper, from outlining the \enquote{Value Alignment Gap} to structuring the three pillars of the CDAVP framework and the illustrative scenarios.
    \item \textbf{Linguistic Refinement:} The AI was used to rephrase complex sentences, improve terminological precision, and enhance the overall clarity and readability of the text.
\end{itemize}

All generated outputs were critically reviewed, validated, and substantially edited by the author. The ultimate intellectual contribution, the core arguments, and the final responsibility for the paper's content lie entirely with the human author.

\bibliographystyle{ACM-Reference-Format} 
\bibliography{references} 

\end{document}